# Architecture and FPGA Implementation of Digital Time-to-Digital Converter for Sensing Applications

Zeinab Hijazi[1], Fatima Bzeih[2], Ali Ibrahim[1]

[1]Department of Electrical and Electronics Engineering, Lebanese International University, Beirut, Lebanon
[2]National Institute for Nuclear Physics (INFN), University of Genova, Genova, Italy
{ali.ibrahim, zeinab.hijazi}@liu.edu.lb

**Abstract.** Many application domains face the challenges of high-power consumption and high computational demands, especially with the advancement in embedded machine learning and edge computing. Designing application-specific circuits is crucial to reducing hardware complexity and power consumption. In these perspectives, this paper presents the design of a Digital Time-to-Digital converter (DTDC) based on multiple delay line topologies. The DTDC is implemented in VHDL for the Xilinx Artix-7 AC701 FPGA device. Simulation results demonstrate the effectiveness of the circuit in converting the input period along a wide range up to 1ps. The designed circuit is implemented with less than 1% of the resource utilization on the target FPGA device.

## 1 Introduction

Time-to-Digital Converters (TDCs) is a crucial need when dealing with time domain signal processing circuits where precise time measurements is required. TDC circuits are essential in many applications such as in time-based ADCs, digital phase locked loops [1], Time Correlated Single Photon Counting [2], telecommunications such as in network timing protocols and high-speed data links, Fluorescence Life Time Measurement [3], LIDAR Systems where TDCs help in measuring the time of flight of laser pulses for distance and speed [4], Diffuse Optical Tomography [5] and medical imaging.

TDCs are block circuits used to measure time intervals (duration) between two successive events usually denoted as a start and stop signal with high precision. The time interval which is defined by the arrival of two signals, the START signal and the STOP signal is then digitized into a numerical value. A simple TDC can be built with a high frequency counter that counts the input time pulse or interval according to a reference clock. Such approach is very simple and may achieve measurements over wide dynamic range, however, measurement errors due the asynchronous nature between the START and STOP signals in this case are significant. Also, this implementation limits the resolution to a clock period of a few nanoseconds. TDCs are classified either to be Analog TDCs or Digital TDCs.

Digital techniques and solutions are becoming more popular due to many advantages that digital circuits can provide compared to their analog counterparts. This is especially evident with technology scaling that led to reduction in area and power consumption, as well as, reduction in gate delays resulting in faster devices. With such techniques, the precision and performance of time measurement of the DTDC can be improved. On the other side, some sensory interfaces deal with wide sensor input range. Such wide sensor range requires to be conditioned and converted to the digital domain for further processing. For example, Metal Oxide based gas sensors (MOX) are one of the sensors dealing with wide input variation. MOX based resistive gas sensors show wide range performance since the baseline resistance depends on several chemical and physical parameters, fabrication process, technology and sensor operating conditions [3]. The MOX sensor in this case is modeled by a variable resistor where its value can range from hundreds of Ohms up to hundreds of Mega or even GOhms. Many MOX based gas sensors uses the resistance based oscillator topology to convert the resistance variation into periodic signal where the obtained output period is proportional to the sensor resistance, $T=\alpha.Rsens$ as shown in Fig. 1. In [3] a wide resistance variation circuit based on resistance-controlled oscillator circuit was proposed. The circuit converts the resistance into a time (period) signal. However, the conversion from time to code is still missing. This work presents the design of a DTDC working on wide range periods, so that, to be converted into a binary code and to be able to tackle the wide sensor resistance range.

The paper is organized as follows: section 2 presents the literature review of DTDCs, section 3 discusses the principle of operation of DTDCs, section 4 describes the circuit implementation, the results are highlighted in section 5, and finally section 6 concludes the paper.

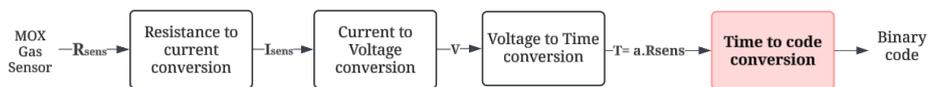

**Fig. 1.** Block diagram of the use of TDC in sensory interface for gas sensing

## 2 Literature review

A DTDC Converter measures the time interval between two events or occurrences. Digital TDCs are mainly based on digital delay lines [6], the most straightforward TDCs measure the time intervals by propagating the START signal in a line of buffers which taps are sampled when the STOP signal arrives, the resultant time resolution is equal to the gate delay in the used technology.

Several DTDC architectures are proposed in literature. Traditional counter based TDCs use high frequency counters which obviously leads to high power consumption. Since higher resolution requires higher operation frequency which in its turn results in additional power consumption increase, Time Interpolation Techniques achieve very high resolution with reduced frequency compared to counter based TDCs of equal resolution [7]. This results in enhanced area and power efficiency of the system.

The conceptual time interpolation technique starts with Tin as input pulse signal for time to pulse generator. The latter is traditionally based on D Flip-Flops and logic gates to provide three signals Tf1 and Tf2 and Tc. Tf1 and Tf2 serve as an input for two fine TDCs or interpolators with finer resolution compared to the clock period (Tclk) while Tc signal is synchronous with the reference clock (clk), thus, ready to be measured by the coarse counter. This architecture suffers from narrow pulse width leading to metastability problem. To avoid such problem, the fractional period widths Tf1/Tf2 of the time-to-pulse generator are extended by one Tclk as proposed in [8]. The state of art here will focus on the improvements/progress in Trapped delay line interpolation technique. A TDC based on trapped delay lines as interpolator is proposed in [9]. The start signal is delayed using buffers before being entered to the latch Flip-flops to be encoded. The output of this topology needs to be converted to a binary code as final output the resolution in theory is equivalent to the propagation delay. The resolution of the TDC is further improved in [10, 11, 12] by replacing the flip-flops with counters. The multiple delay line in this architecture quantizes the input signal Tin by multiple delays of the system clock. Summing up the results of all counters will provide the output as a code to be converted to binary. The equivalent resolution in this case is equal to the cell delay or the phase shift among clocks. This paper adopts the later DTDC architecture based on multiple delay lines.

To minimize the resolution even more i.e. improve the effective TDC resolution, it is possible to increase the wrapping density of the delay clocks through increasing the number of delay cells in the delay line [10]. In [9] and [13], a method applied to parallel measurements with multiple delay cores is proposed. The delay cells are combined and merged to create homogenous delays $\Delta\tau$. As a result, such multiple merged delay cells are placed as a new cell in the delay line to form nearly uniform cell delay.

## 3   DTDC Principle of Operation

DTDC based on multiple delay lines [7] consists of the following blocks: Time-To-Pulse Generator, two fine TDCs, clock source, Coarse Counter and ALU. Fig. 2 shows the block diagram of the DTDC architecture. Each of the Fine TDCs (Fine TDC 1 and Fine TDC 2) consists of Multiple Delay Lines, Counter Array and a Summer/Adder. The role of each block is as follows:

The input pulse Tin of the Time-To-Pulse Generator is segmented into three pulses Tf1, Tf2 and Tc to feed the inputs of the coarse TDC and the multiple delay lines of the fine TDC1 and TDC2. A clock source feeds the counter array in fine TDC1, fine TDC2, coarse TDC/Counter and Time-To-Pulse Generator. The Coarse Counter/TDC endures an approximation of time measurements with lower resolution and higher operation speed than the fine TDC to support the overall system in an accurate time measurement. The final time measurements occur inside the ALU and fed at the output "Dout" after taking into consideration and combining the final measured time interval resulted from coarse Counter/TDC, fine TDC1 and fine TDC2.

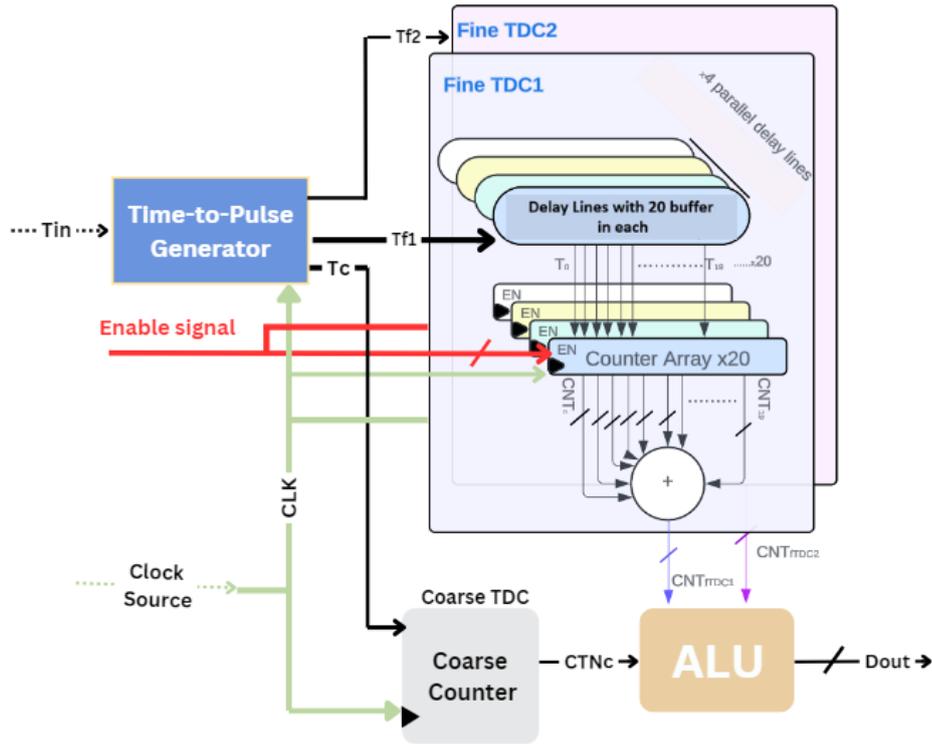

**Fig. 2.** Block diagram of the DTDC architecture.

## 4 Circuit Implementation

A clock source feeds the Time-To-Pulse Generator and the TDC Counters. The Time-to-Pulse generator is based on Nutt Interpolation Technique with extended fractional period widths to avoid narrow pulse widths. So, Tf1 and Tf2 are extended by one Tclk. The Time-To-Pulse Generator consists of D flip-flops, NOR gate, XOR gate and NOT gate, where the output Tf1 and Tf2 are the output of the two NOR gate, and the output Tc is the output of the XOR gate [7]. The interval of the input pulse Tin can be calculated using the equation 1:

$$Tin = Tc + Tf1 - Tf2 \qquad (1)$$

For the two fine TDCs, this work adopts the parallel Multiple Delay Line (MDL) topology [7]. The number of the MDL can include many delay lines in parallel depending on the targeted resolution. The MDL in this work consists of 4 parallel delay lines for both fine TDC1 and TDC2. All the delay lines have the same one-shot pulse input coming from the Time-To-Pulse Generator (TPG) output with Tf1 for Fine TDC1 and Tf2 for Fine TDC2. Each delay line in its turn consists of 20 buffers or delay steps. The role of each buffer is to act exactly as a delay element, the one-shot pulse is propagated in each delay line, and at each buffer. The output of each buffer in the same delay line is connected to the input 'In' of each counter in the same counter-array. Since each counter is specified for only one output buffer then 20 counters in each counter-array are needed. All the counters in the same counter-array have the same enable input bit and are triggered at the same time, whereas, all counter-arrays are triggered by order and not at the same time. When a specific counter-array is triggered, all its outputs value is directed to the adder circuit. In the proposed design, each counter-array needs a total of 19 adders, each adder has two 35 bits input and one 35 bits output. After the combination of the four "19 adders", a Multiplexer is used to select the output of the last adder that corresponds to a specific counter-array assigned to the Fine TDC output.

The role of the coarse counter is when the input Tc is rising, active high or low, the counter counts the number of rising edges that occurs during this variation of amplitude which result in finding the count value of the input Tin with a delayed time. When the one-shot input pulse Tc is zero after all its transitions, the counting value of the Coarse-Counter is different than zero, thus, the counting value is assigned to the output. The ALU then performs Arithmetic and Logic operations, since the input Tin is equal to (Tc + Tf1 – Tf2), the ALU adds the output of the Coarse Counter to the output of Fine TDC1 then subtract the addition by the output of Fine TDC2, in this case the result is the counting value of the primary input Tin.

## 5  Results

The circuit is designed in VHSIC hardware description language (VHDL) at register transfer level (RTL). The circuit is simulated and implemented on Vivado tool for Xilinx FPGA devices. The clock input is set to 800 MHz achieving an output range up to 1ps. The maximum achieved time (1/Tclk) is constrained by the proposed architecture itself and the number of used buffers and delay lines in the fine TDC block. While the minimum time depends on the input clock source since the time to pulse generator is sourced by this main clock. Fig. 3 illustrates the simulation results for a Tin = 100 ns. As clearly shown in the figure, the time interval of the input signal Tin is 100 ns between 830 ns and 930 ns. The three signals created by the time to pulse generator namely tf1, tf2, and tc are shown at the bottom of the figure to highlight the correct implementation of the time to pulse generation. The DTDC output is reporting the number 100 which is coherent with the input signal interval demonstrating the correct functionality of the proposed circuit.

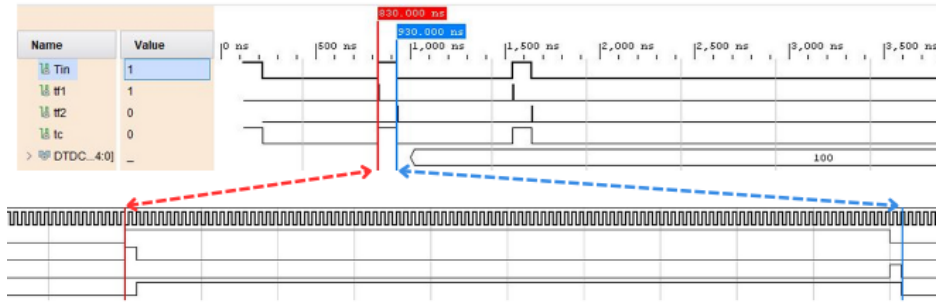
**Fig. 3.** Behavioural simulation of the DTDC for a Tin = 100 ns.

Tables 1 reports the resource utilization for the proposed DTDC when implemented on Artix-7 AC701 Xilinx FPGA device.

**Table 1.** FPGA Resources Utilization

|                 | DTDC |
|-----------------|------|
| Slice LUT       | 177  |
| Slice Registers | 249  |
| Bounded IOB     | 42   |
| BUFGCTRL        | 3    |

## 6 Conclusion

This paper presented the design of DTDC based on multiple delay line architecture. The architecture deployed 2 fine TDC architecture with 4 parallel delay lines each with 20 delay buffers and 4 counter arrays each of 20 counters reading from the 20 buffers. The DTDC is implemented in VHDL for the Xilinx Artix-7 AC701 FPGA device. Simulation results demonstrated the correctness of the circuit in converting input period into digital codes. The designed circuit is implemented with less than 1% of the resource utilization on the target FPGA device. Future work will consist of increasing the measurement range of the TDC trying not to affect its hardware resources. Moreover, a reconfigurable delay line architecture would be of interest enable the DTDC to be employed with different applications.

**Acknowledgments**
This project has received funding from the European Union's Horizon Europe research and innovation program under the Marie Sklodowska-Curie grant agreement No 101086359.